# From in silico to in vitro: A comprehensive review of validating bioinformatics findings

**Tian-Yang Wang[1], Si-Lin Chen[2], Yun-Ze Wang[3], Yi-Chao Zhang[4], Xin-Yuan Song[5], Zi-Qian Bi[6], Ming Liu[7], Qian Niu[8], Jun-Yu Liu[8], Poh-Sun Feng[9], Xin-Tian Sun[10], Ben-Ji Peng[11, 12], Cheng Fei[13], Lawrence KQ Yan[14], Ri-Yang Bao[5], Zi-Yuan Qin[5], Ze-Kun Jiang[15*]**

[1]Department of Computer Science, University of Liverpool, Liverpool L69 7ZX, England, United Kingdom. [2]College of Computer Science and Technology, Zhejiang University, Hangzhou 310058, Zhejiang, China. [3]School of Informatics, University of Edinburgh, Edinburgh EH8 9YL, Scotland, United Kingdom. [4]Department of Computer Science, The University of Texas at Dallas, Richardson 75080-3021, Texas, USA. [5]Department of Computer Science, Emory University, Atlanta 30322, Georgia, USA. [6]Department of Computer Science, Indiana University, Bloomington 47405, Indiana, USA. [7]Department of Computer Science, Purdue University, West Lafayette 47907, Indiana, USA. [8]Graduate School of Informatics, Kyoto University, Kyoto 606-8501, Kyoto, Japan. [9]Department of Computer Science and Information Engineering, National Taiwan Normal University, Taipei 10610, Taiwan, China. [10]School of Computing Science, Simon Fraser University, Burnaby V5A 1S6, British Columbia, Canada. [11]School of Computer Science, Georgia Institute of Technology, Atlanta 30332, Georgia, USA. [12]AppCubic, Miami, Florida, USA. [13]Department of Computer Sciences, University of Wisconsin-Madison, Madison 53706, Wisconsin, USA. [14]Department of Computer Science and Engineering, The Hong Kong University of Science and Technology, Clear Water Bay, Kowloon, Hong Kong, China. [15]West China Biomedical Big Data Center, West China Hospital, Sichuan University, Chengdu 610041, Sichuan, China.

*Correspondence to:** Ze-Kun Jiang. West China Biomedical Big Data Center, West China Hospital, Sichuan University, No. 37, Guoxue Alley, Chengdu 610041, China. E-mail: zekun_jiang@163.com.

***Competing interests***

The authors declare no conflicts of interest.

***Acknowledgments***

The authors would like to acknowledge all contributors to this research and the institutions that supported this work. The study was supported by the Key Research Project of Science & Technology Department of Sichuan Province, China (2024YFFK0041) and the Science and Technology Program of Xinjiang Uyghur Autonomous Region (2025E01032).

***Abbreviations***

CETSA, cellular thermal shift assay; Co-IP, co-immunoprecipitation; CRISPR, clustered regularly interspaced short palindromic repeats; DEGs, differentially expressed genes; ELISA, enzyme-linked immunosorbent assay; FDR, false discovery rate; FISH, fluorescence in situ hybridization; FRET, fluorescence resonance energy transfer; GEO, Gene Expression Omnibus; GO, Gene Ontology; GSEA, gene set enrichment analysis; IHC, immunohistochemistry; ISH, in situ hybridization; KEGG, Kyoto Encyclopedia of Genes and Genomes; MST, microscale thermophoresis; PCA, protein fragment complementation assay; PLA, proximity ligation assay; PPI, protein–protein interaction; qPCR, quantitative polymerase chain reaction; RNA-seq, RNA sequencing; RT-PCR, reverse transcription polymerase chain reaction; SPR, surface plasmon resonance; TCGA, The Cancer Genome Atlas; Y2H, yeast two-hybrid.

***Citation***

Wang TY, Chen SL, Wang YZ, et al. From in silico to in vitro: A comprehensive review of validating bioinformatics findings. *Med Data Min*. 2026;9(3):18. doi: 10.53388/MDM202609018.

***Peer review information***

*Medical Data Mining* thanks Zheng-Rui Li and other anonymous reviewers for their contribution to the peer review of this paper.

**Editorial Advisory Board:** Jing-Bo Zhai.

**Production editor:** Ming-Hao Wang.

## Abstract

Translating computational predictions into experimentally validated biological knowledge remains one of the central challenges in modern bioinformatics. This review critically examines how in silico findings can be prioritized, tested, and interpreted through experimental validation. We organize the validation process around three recurring challenges: the specificity gap between genome-wide computational predictions and single-target experimental assays; the reproducibility-translatability tension, in which results validated in one model system may not generalize to another; and the scale-depth trade-off between high-throughput candidate discovery and the low-throughput nature of mechanistic validation. Rather than presenting an encyclopedic catalogue of techniques, we compare the strengths, limitations, and common failure modes of major validation approaches, including qPCR, RNA-seq, Western blotting, co-immunoprecipitation, luciferase reporter assays, CRISPR perturbation, and functional phenotypic assays. We also provide structured comparison tables for gene expression, protein–protein interaction, non-coding RNA, regulatory element, and pathway validation, together with decision-making frameworks to guide method selection according to prediction type, biological context, evidence stringency, throughput, and resource constraints. Case studies from cancer genomics, drug target discovery, miRNA regulation, and neurological disease illustrate how multi-step validation workflows can strengthen causal inference and reduce false-positive interpretation. Finally, we discuss how emerging technologies, including CRISPR screens, single-cell and spatial multi-omics, and AI-assisted experimental design, may reshape validation practice by improving scalability, context specificity, and reproducibility.

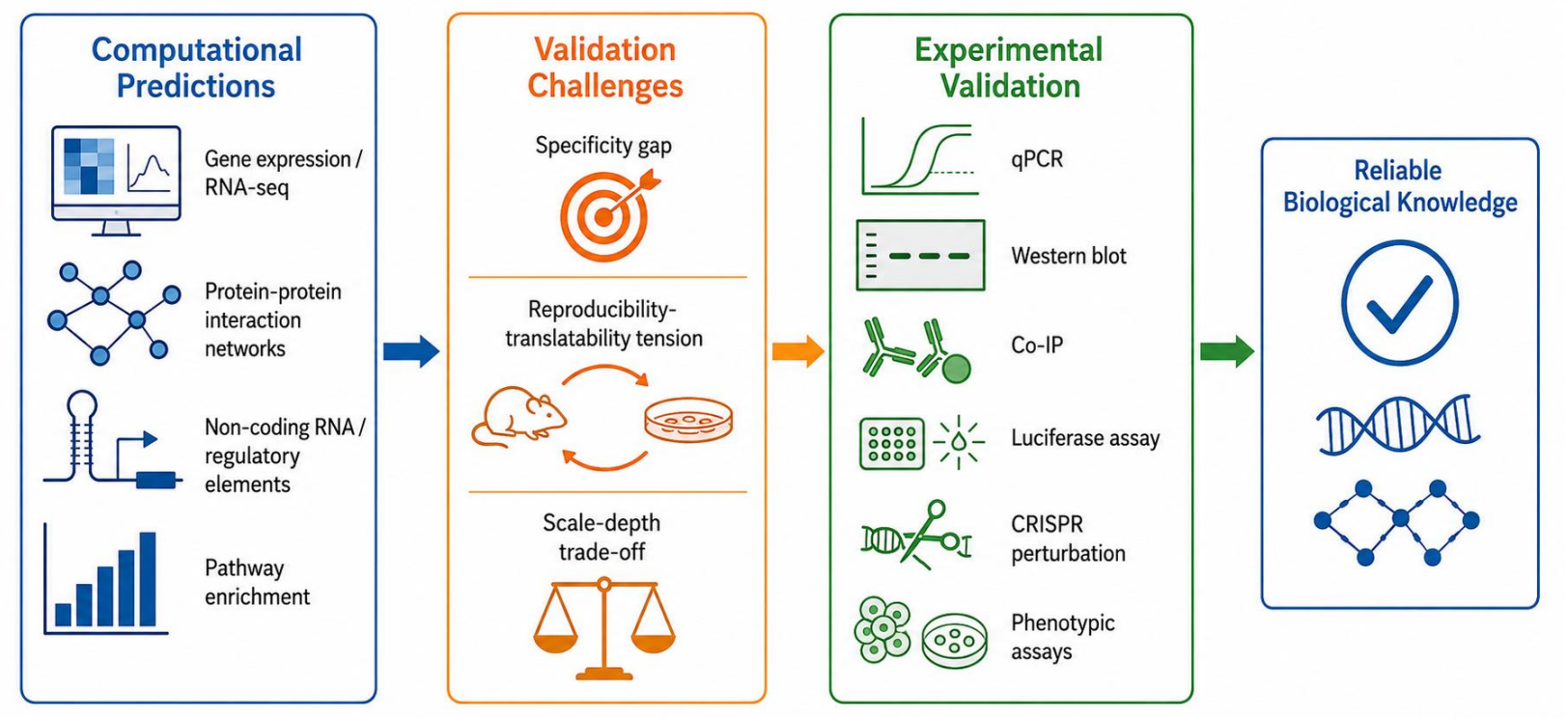

Graphic abstract.

## Introduction

Bioinformatics has transformed biological research by enabling the computational analysis of genome-scale datasets, from whole-genome sequencing and transcriptomics to proteomics and epigenomics [1, 2]. The exponential growth of high-throughput data generation has produced an unprecedented volume of computational predictions regarding gene function, protein interactions, regulatory networks, and disease mechanisms [3, 4]. Yet, the biological community has increasingly recognized that the rate at which predictions are generated far outstrips the capacity to validate them experimentally, creating a widening gap between computational hypothesis generation and empirical confirmation [5].

This gap is not merely quantitative but is shaped by three distinct and interrelated challenges. First, the *specificity gap*: computational methods operate at a genome-wide scale, identifying hundreds or thousands of candidate genes, interactions, or regulatory elements simultaneously, whereas experimental validation typically proceeds one gene, one interaction, or one regulatory element at a time [6, 7]. A differential expression analysis may identify many significantly altered genes, whereas qPCR validation is typically feasible only for a selected subset. Second, the *reproducibility–translatability tension*: a finding validated in one cell line, one organism, or one experimental system may not reproduce in another [5, 8]. A protein–protein interaction confirmed by co-immunoprecipitation in HEK293T cells may fail to replicate in primary neurons, not because the original finding was wrong, but because the interaction is context-dependent. Third, the *scale–depth trade-off*: high-throughput validation methods (such as CRISPR screens) sacrifice depth and mechanistic insight, while gold-standard low-throughput methods (such as co-immunoprecipitation followed by Western blot) provide mechanistic detail but cannot scale to validate the full output of a computational analysis [9, 10].

Importantly, computational analyses in bioinformatics serve fundamentally different purposes, and not all outputs require—or are intended for—direct experimental validation in the traditional sense. It is useful to distinguish three stages of computational investigation: (1) *discovery*, in which systems-level analyses (e.g., pan-cancer multi-omics integration) identify candidate gene families, pathways, or regulatory networks and set the conceptual direction for a field [11, 12]; (2) *prioritization*, in which focused analyses across independent cohorts narrow the candidate space to specific genes or biomarkers suitable for further investigation [13, 14]; and (3) *mechanistic validation*, in which targeted experiments (qPCR, Western blot, CRISPR perturbation, functional assays) establish causal or mechanistic evidence for individual candidates [15]. These stages do not occur simultaneously and may span multiple studies over years—for example, in the voltage-gated sodium channel (VGSC) field, pan-cancer characterization of the gene family preceded biomarker-level prioritization of SCN3B in glioma, which in turn preceded functional validation of its role in cell motility [11, 13, 15]. Discovery-stage studies may require only internal consistency and independent cohort replication; prioritization-stage studies benefit from cross-dataset confirmation and preliminary functional evidence; only at the mechanistic validation stage is comprehensive multi-method experimental verification essential. This review focuses primarily on the third stage while clarifying where earlier-stage analyses may appropriately conclude without bench-level experiments.

Existing reviews of bioinformatics validation have tended to catalog experimental methods encyclopedically, describing what each technique does without critically evaluating when it should or should not be used, what its common failure modes are, or how to choose between competing approaches [16]. While such reviews serve as useful reference documents, they provide insufficient guidance for researchers facing the practical challenge of designing a validation strategy for a specific set of computational predictions. The absence of decision-making frameworks, structured comparison tables, and critical analysis of method limitations has left a significant gap in the literature [17]. While best-practice guidelines exist for individual analytical steps—such as RNA-Seq analysis [18] and single-cell data processing [19]—an integrated framework guiding the complete computational-to-experimental validation workflow remains lacking.

This review addresses that gap by providing a critical, comparative analysis of the validation process. For each major category of bioinformatics prediction—gene expression, protein–protein interactions, non-coding RNA targets, and pathway enrichment—we evaluate the available validation methods with respect to their sensitivity, specificity, throughput, cost, and common error modes. We present structured comparison tables that enable researchers to select appropriate validation strategies based on their specific experimental context and resource constraints. We complement this analytical framework with detailed case studies that illustrate complete validation workflows, from initial computational prediction through candidate prioritization to multi-method experimental confirmation, highlighting both successes and failures.

This review is intended for both computational biologists seeking to design rigorous validation experiments for their predictions and experimental biologists evaluating computational results for downstream functional testing. Rather than providing an encyclopedic catalog, we emphasize when and why to use specific approaches, highlight common failure modes, and offer practical decision-making frameworks. The overall validation workflow is summarized in Figure 1.

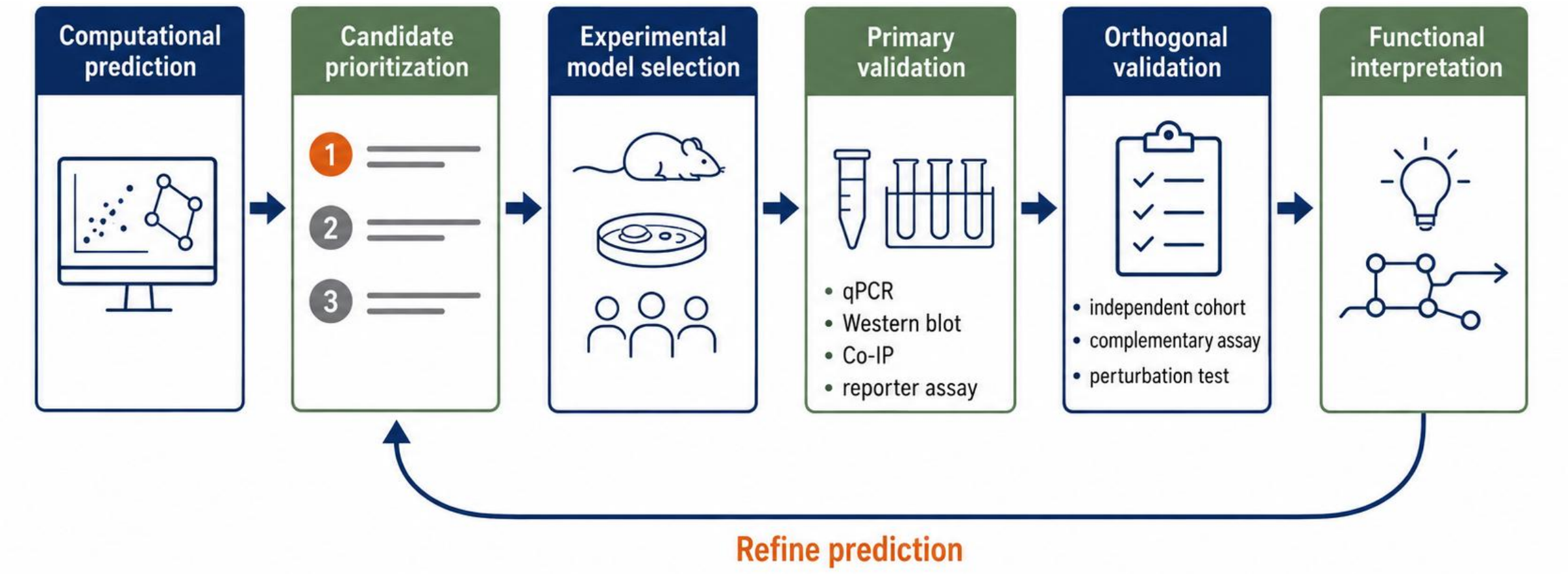

**Figure 1 Computational-to-experimental validation workflow for bioinformatics findings.**
Computational predictions are first prioritized, then tested in context-matched experimental systems using primary and orthogonal validation methods, with discordant findings feeding back into prediction refinement. Co-IP, co-immunoprecipitation; qPCR, quantitative polymerase chain reaction.

## Scope and literature search methodology

This is a narrative review. The literature search was conducted across PubMed, Web of Science, Scopus, and Google Scholar using combinations of the following primary search terms: (bioinformatics OR computational biology OR in silico) AND (experimental validation OR wet-lab validation OR in vitro validation OR functional validation). Additional topic-specific search terms were used for individual sections (e.g., "qPCR validation," "co-immunoprecipitation," "miRNA target validation," "CRISPR screen," "Perturb-seq," "single-cell validation"). No date restrictions were applied; however, we prioritized publications from 2015–2025 while retaining foundational references that established key methods or concepts. Inclusion criteria required that articles address the experimental validation of computationally derived predictions, encompassing original research, method comparisons, and reviews. Articles reporting purely computational analyses without experimental follow-up, purely experimental studies without computational prediction components, and conference abstracts were excluded. Reference lists of included articles and relevant reviews were screened to identify additional pertinent studies (snowball approach). The literature search was last updated in January 2025.

As a narrative review, this work is subject to several inherent limitations that should be acknowledged. *Publication bias* is a primary concern: studies reporting successful validation of computational predictions are more likely to be published than those reporting negative or inconclusive results, which may lead to an overestimation of the concordance rates between computational predictions and experimental outcomes discussed herein. *Language bias* may also be present, as the search was restricted to English-language publications, potentially excluding relevant studies published in other languages. *Selection bias* in the literature itself—whereby well-resourced model organisms and disease contexts (e.g., cancer, neurological disease) are disproportionately represented relative to less-studied biological systems—may limit the generalizability of some methodological recommendations. We have endeavored to mitigate these biases by including studies reporting null results where available, by drawing on multiple independent databases, and by explicitly noting where evidence bases are sparse or unevenly distributed across biological contexts.

## Computational predictions and their validation requirements

Bioinformatics encompasses the computational analysis of diverse biological data types—genomic, transcriptomic, proteomic, and epigenomic—each of which presents distinct validation challenges [20, 21]. Genomic data, including whole-genome and exome sequencing, supports variant calling, copy number variation analysis, and mutation–phenotype association studies [22]. Transcriptomic data, primarily generated through RNA-Seq, enables differential gene expression analysis, alternative splicing detection, and co-expression network construction [23, 24]. Proteomic data, obtained through mass spectrometry, provides information on protein abundance, post-translational modifications, and protein complexes [25, 26]. Epigenomic data captures regulatory modifications such as DNA methylation patterns and histone marks via techniques including bisulfite sequencing and ChIP-Seq [27, 28]. The nature of the underlying data fundamentally determines both the type of computational prediction generated and the experimental approach required for its validation. Importantly, recent benchmarking efforts have revealed that the choice of computational workflow—including normalization, alignment, and statistical testing—can substantially influence which predictions emerge as candidates for validation, underscoring the need for best-practice guidelines at each stage of the analytical pipeline [18, 29]. Critically, the purpose of a computational analysis determines the level of experimental validation that is appropriate. Systems-level integrative analyses—such as pan-cancer multi-omics characterization of a gene family [11] or comprehensive oncogenic determinant mapping [12]—serve a *discovery* function, and their primary contribution lies in framework construction and hypothesis generation rather than in individual gene-level proof. Conversely, target-directed analyses that nominate specific biomarkers or therapeutic candidates require progressively more stringent validation, culminating in functional experiments when mechanistic claims are made. The validation strategies discussed in subsequent sections should therefore be applied with awareness of this distinction: not every computational output warrants the same depth of experimental follow-up.

### Core analytical methods and their limitations

The principal analytical methods in bioinformatics each carry characteristic strengths and weaknesses that must be understood before designing validation experiments. *Differential expression analysis* uses statistical frameworks (e.g., DESeq2, edgeR, limma) to identify genes whose expression differs significantly between conditions [30–32]. However, these methods are sensitive to normalization strategy, batch effects, biological replication, and multiple-testing thresholds; these analytical choices can substantially alter which candidates are forwarded for validation [33].

*Protein–protein interaction (PPI) prediction* methods—including structure-based docking, sequence-based homology inference, network-based approaches, and deep sequence-based models—generate interaction predictions with widely varying confidence levels [34–37]. A persistent limitation is that computational PPI predictions aggregate interactions across tissues, developmental stages, and cellular contexts, whereas most interactions are highly context-dependent. The STRING database, for instance, assigns confidence scores that aggregate text-mining, co-expression, and experimental evidence, but a high STRING score does not guarantee that two proteins interact in a specific cell type of interest [38].

*Pathway and network analysis*, including GSEA, over-representation analysis, and network topology methods, identifies biological processes enriched in a gene list [39–41]. Common pitfalls include gene-set redundancy (many GO terms overlap substantially), sensitivity to background gene list selection, and the ecological fallacy of assuming that enrichment of a pathway implies coordinated activation of all constituent genes [42].

*Machine learning and predictive modeling*, including support vector machines, random forests, and deep neural networks, are increasingly used for gene function prediction, variant pathogenicity scoring, and drug response prediction [43–46]. The primary validation concern is overfitting: models trained and evaluated on the same or highly correlated datasets may show inflated performance metrics that do not generalize to independent datasets.

### Prediction types and validation requirements

Table 1 provides a structured overview of the major bioinformatics prediction types, their associated computational methods, common error modes, recommended validation approaches, and the stringency of validation required. This table is intended to serve as a practical reference for researchers designing validation experiments.

### From computational predictions to wet-lab validation

In practice, validation should proceed from prediction type identification to model selection, primary validation, orthogonal confirmation, and iterative refinement when results are discordant.

The transition from computational predictions to wet-lab validation requires systematic decision-making regarding the selection of experimental models and validation methods [47]. This process typically begins with hypothesis formulation based on computational analysis, proceeds through experimental model selection, and culminates in iterative testing and refinement of predictions [48, 49]. The choice of experimental system fundamentally influences both the feasibility and the interpretability of validation results, and mismatches between the computational prediction context and the validation system are a major source of irreproducible findings.

**Experimental model systems: A comparative assessment**

Researchers must select from among cell culture models, animal models, and organoid systems, each of which offers distinct advantages and imposes characteristic limitations. Table 2 summarizes these trade-offs to facilitate informed model selection.

Cell culture models, including immortalized cell lines and primary cultures, are widely used for initial validation because of their low cost, high throughput, and experimental tractability. They enable rapid genetic manipulation via transfection, transduction, or CRISPR editing, and are suitable for high-throughput screening applications. However, two-dimensional cell culture systems lack the tissue architecture, cellular heterogeneity, and microenvironmental cues present in vivo, which can profoundly affect gene expression, protein interactions, and pathway activity. Furthermore, commonly used immortalized lines (e.g., HeLa, HEK293T) have undergone extensive genomic rearrangements that may render them poor models for normal tissue biology [50].

Animal models, particularly genetically modified mice, zebrafish, and Drosophila, provide physiological complexity and systemic context that are absent from cell culture. However, their translatability is limited by species-specific biology. For instance, although most mouse genes have human counterparts, not all have simple one-to-one orthology, and species-specific gene family expansion or functional divergence can complicate extrapolation from mouse models to humans [51]. In addition, drug-metabolizing enzymes, particularly CYP isoforms, differ substantially between rodents and humans in expression, catalytic activity, induction, and inhibition [52].

Organoid and three-dimensional culture models represent an intermediate approach that recapitulates key features of tissue architecture while retaining the experimental tractability of in vitro systems [53–56]. Patient-derived organoids, in particular, enable personalized validation of computational predictions in a genetically relevant context. However, organoid systems remain technically demanding, exhibit considerable batch-to-batch variability, and lack vascularization and immune components.

**Table 1 Major bioinformatics prediction types and recommended validation strategies**

| Prediction type | Computational methods | Common error modes | Recommended validation | Validation stringency |
|---|---|---|---|---|
| Gene Expression (DEGs) | DESeq2, edgeR, limma; fold-change filtering | Batch effects; normalization artifacts; low replicate number; FDR threshold sensitivity | qPCR (primary); Western blot (protein-level); RNA-Seq (independent cohort) | Moderate: ≥ 2 independent methods |
| Protein–Protein Interaction | STRING, BioGRID, molecular docking, deep learning | Context-dependent interactions reported as universal; text-mining false positives; homology-inferred artifacts | Co-IP + Western blot (primary); Y2H or PCA (orthogonal); FRET (in vivo) | High: ≥ 2 orthogonal methods required |
| miRNA Targets | TargetScan, miRDB, miRanda; seed-match algorithms | Seed-match alone insufficient; thermodynamic stability overestimated; tissue-specific expression ignored | Dual-luciferase assay (direct binding); qPCR + Western blot (functional effect); Ago-CLIP (transcriptome-wide) | High: direct binding + functional evidence |
| Pathway Enrichment | GSEA, DAVID, Reactome, KEGG pathway analysis | Gene-set redundancy; background list bias; ecological fallacy; annotation incompleteness | Pharmacological inhibition/activation; reporter assays; multi-gene perturbation (CRISPR) | Moderate–High: functional perturbation preferred |
| Regulatory Elements | ENCODE, motif scanning, ChIP-Seq peak calling | Motif occurrence functional binding; cell-type specificity; ChIP-Seq antibody quality | ChIP-qPCR (binding confirmation); luciferase reporter (functional activity); CRISPR deletion (necessity) | High: binding + functional activity |

**Table 2 Experimental model systems for validating bioinformatics predictions**

| Model type | Complexity | Cost | Throughput | Physiological relevance | Key limitations |
|---|---|---|---|---|---|
| 2D Cell Culture | Low | Low | High | Low–Moderate | Lacks tissue architecture; genomic drift in immortalized lines; no microenvironment |
| Primary Cell Culture | Moderate | Moderate | Moderate | Moderate | Limited passage number; donor variability; loss of tissue context |
| Animal Models (mouse, zebrafish) | High | High | Low | High | Species differences; ethical constraints; long timelines; limited human translatability |
| Organoids / 3D Culture | Moderate–High | Moderate–High | Low–Moderate | Moderate–High | Batch variability; no vascularization; no immune component; technical expertise required |
| Patient-Derived Xenografts | Very High | Very High | Very Low | High | Extremely resource-intensive; mouse stroma replacement; limited scalability |

## Choosing the right experimental methods

After an appropriate experimental model is selected, validation methods should be matched to the specific claim being tested. Expression predictions generally require transcript-level confirmation followed by protein or functional assays when mechanistic claims are made; predicted PPIs require biochemical or cell-based interaction assays; and gene-function predictions require perturbation with appropriate negative, rescue, and orthogonal controls. Concordance across independent methods provides stronger evidence than repeated use of a single assay.

### Gene expression prediction validation

Gene expression prediction validation confirms whether computationally nominated expression changes are reproducible and biologically meaningful. Method selection should be guided by the question being asked, the number of candidates, the need for quantitative precision, and whether spatial or functional evidence is required.

Quantitative PCR (qPCR) and reverse transcription PCR (RT-PCR) remain useful for targeted confirmation of individual transcript-level predictions. Their reliability depends on primer specificity, stable reference genes, independent biological replication, and transparent reporting of assay quality; because qPCR measures RNA only, protein-level or functional assays are needed when mechanistic claims are made.

Independent RNA-Seq can validate larger candidate sets and test whether discovery-stage expression patterns reproduce in a separately generated dataset. True validation requires an independent biological cohort or independently processed samples; reusing the discovery dataset inflates apparent concordance and does not provide independent confirmation.

Western blotting and ELISA test whether transcript-level predictions are reflected at the protein level. Their reliability depends on validated antibodies, appropriate controls, linear-range quantification, and transparent reporting of assay validation or unprocessed blot data.

Immunohistochemistry (IHC) and in situ hybridization (ISH) add spatial context to homogenate-based assays and are useful when tissue localization or cell-type-specific expression is central to the prediction. Interpretation should account for antibody or probe specificity, fixation artifacts, and scoring variability; quantitative image analysis can reduce subjectivity.

Gene overexpression, knockdown, and knockout models provide functional evidence that complements expression profiling. Gain- and loss-of-function experiments are strongest when supported by multiple independent reagents, rescue experiments, and controls for off-target effects, compensation, and overexpression artifacts.

## Protein–protein interaction validation

Validating protein–protein interactions (PPIs) is essential for confirming computationally nominated molecular networks. Because PPI predictions vary in confidence and biological context, experimental confirmation should be chosen according to whether the question concerns co-complex membership, direct binary interaction, subcellular proximity, or functional consequence.

The selection of a PPI validation method should be driven by the specific question being addressed. If the goal is to confirm that two proteins exist in the same complex in a physiologically relevant context, Co-IP followed by Western blot or mass spectrometry is appropriate. If the question concerns direct binary interaction, yeast two-hybrid or FRET provides more relevant evidence. If the goal is to map the complete interactome of a bait protein, affinity purification–mass spectrometry (AP-MS) is the method of choice. For high-confidence validation, two orthogonal methods should be employed—for example, Co-IP to demonstrate co-complex formation and FRET to confirm proximity in living cells. Interactions that are confirmed by only a single method, particularly a method with a high false-positive rate (e.g., yeast two-hybrid alone), should be interpreted with caution.

### Co-immunoprecipitation (Co-IP) studies

Co-immunoprecipitation (Co-IP) tests whether proteins exist in the same complex under near-native lysis conditions, but it cannot distinguish direct from bridged interactions and is sensitive to antibody specificity and lysis or wash conditions. IgG, input, and knockout or knockdown controls are essential for interpretation.

### Yeast two-hybrid screening

Yeast two-hybrid (Y2H) screening detects binary interactions and is useful for high-throughput discovery, but hits require orthogonal validation because the yeast nuclear context, missing post-translational modifications, self-activating bait proteins, and membrane-protein incompatibility can produce false positives or false negatives.

### Fluorescence resonance energy transfer (FRET)

Fluorescence resonance energy transfer (FRET) detects protein proximity in living cells and can provide spatial and temporal information about candidate interactions. Its main limitations are fluorescent tagging, overexpression artifacts, and orientation-dependent signal efficiency, so expression-level controls and donor/acceptor controls are required.

### Mass spectrometry-based protein interaction mapping

Affinity purification followed by mass spectrometry (AP-MS) identifies proteins that co-purify with a bait protein and is useful for mapping protein complexes. Quantitative controls and contaminant-scoring methods are needed because abundant non-specific binders can obscure true interaction partners.

### Protein fragment complementation assay (PCA)

Protein fragment complementation assays detect interactions by reconstituting a split reporter when two candidate proteins are brought into proximity. They are useful in living cells but require controls for forced proximity, tag effects, reporter stability, and irreversible complementation in BiFC-style systems.

## Validation of non-coding RNAs and regulatory elements

Non-coding RNAs (ncRNAs) and regulatory elements play crucial roles in gene regulation and cellular processes. The computational prediction of ncRNA targets and regulatory element function is particularly challenging because these predictions depend on sequence features (e.g., miRNA seed matches, transcription factor binding motifs) that are necessary but not sufficient for biological function. This section discusses experimental techniques for validating these predictions and highlights common pitfalls.

### miRNA target prediction and validation

MicroRNAs (miRNAs) regulate gene expression by binding to complementary sequences in messenger RNA 3’ untranslated regions (UTRs), leading to translational repression or mRNA degradation [57]. Computational miRNA target prediction generally relies on sequence complementarity, seed-region matching, conservation, and model-specific features, while curated experimentally validated interactions can be cross-checked against resources such as miRTarBase [58]. Experimental validation typically employs luciferase reporter assays, in which the predicted target 3’ UTR is cloned downstream of a luciferase reporter [59]. If the miRNA binds the 3’ UTR, luciferase activity is suppressed, indicating a direct functional interaction [60]. This is complemented by qPCR and Western blot measurements of endogenous target expression following miRNA overexpression or inhibition.

A persistent challenge is the low concordance between different target prediction algorithms, which frequently disagree on specific targets. Computational miRNA target predictions have highly variable positive predictive value, and one expression-based evaluation found that the proportion of predicted targets supported by downstream expression evidence ranged from 7.83% to 25.68%, depending on the analytical model used [61]. Luciferase reporter assays remain useful for direct binding tests, but they are performed outside the endogenous genomic context; therefore, direct binding evidence should be paired with endogenous expression and functional evidence before a miRNA-target interaction is considered validated.

### ChIP-Seq for transcription factor binding site analysis

Chromatin immunoprecipitation followed by sequencing (ChIP-Seq) is a powerful technique for genome-wide identification and validation of transcription factor binding sites [62]. ChIP-Seq provides high-resolution binding profiles that can confirm computationally predicted binding sites and map regulatory regions controlling gene expression. The technique involves crosslinking proteins to DNA, immunoprecipitating the protein of interest, and sequencing the associated DNA fragments.

Critical limitations include dependence on antibody quality: ChIP-grade antibodies must be rigorously validated, as non-specific antibodies generate misleading binding profiles. Peak-calling algorithms can produce false positives, particularly in regions of high chromatin accessibility or repetitive sequences. Furthermore, ChIP-Seq identifies binding sites but does not directly demonstrate functional regulation; a transcription factor may bind a genomic region without affecting the expression of nearby genes. Combining ChIP-Seq with functional assays (e.g., reporter gene assays or CRISPR-mediated deletion of the binding site) is necessary to establish functional significance. CUT&RUN and CUT&Tag are emerging alternatives that require fewer cells and generate lower backgrounds than traditional ChIP-Seq [63].

### Functional experiments for long non-coding RNAs (lncRNAs)

Long non-coding RNAs (lncRNAs) participate in diverse regulatory processes, including chromatin remodeling, transcription regulation, and post-transcriptional RNA processing [64]. Functional validation of lncRNAs typically involves loss-of-function experiments using RNAi, antisense oligonucleotides (ASOs), or CRISPR-mediated deletion [65]. Gain-of-function experiments through ectopic expression can complement loss-of-function data [66]. RNA pull-down assays and RNA immunoprecipitation (RIP) identify protein binding partners, providing mechanistic insight [67].

LncRNA validation presents unique challenges. Many lncRNAs function in cis (regulating neighboring genes), and CRISPR deletion of the lncRNA locus may inadvertently disrupt neighboring regulatory elements, confounding interpretation. Conversely, RNAi primarily targets cytoplasmic RNAs and may be ineffective against nuclear-localized lncRNAs. ASOs provide an alternative for nuclear targets but have their own off-target effects. Overexpression from a transgene may not recapitulate cis-regulatory functions. Rescue experiments, in which the lncRNA is re-expressed from an ectopic locus after deletion, are essential but may fail for lncRNAs that function in cis. These complexities demand careful experimental design and cautious interpretation of results [68].

### Reporter gene assays for regulatory element function validation

Reporter gene assays assess the functional activity of regulatory elements by placing a reporter gene (e.g., luciferase, GFP, or β-galactosidase) under the control of a candidate regulatory element [69–71]. When the regulatory element is active, it drives reporter expression, which is then quantified [72, 73]. This approach is widely used to validate predicted enhancers, promoters, and silencers identified through bioinformatics tools.

Reporter assays test regulatory element activity outside its native genomic context, which can produce both false positives (elements that are active in the reporter construct but not at their endogenous locus) and false negatives (elements whose activity depends on chromatin context or long-range interactions not recapitulated in the reporter). Additionally, the commonly used practice of testing isolated fragments may miss regulatory elements whose activity depends on combinatorial interactions with neighboring elements. Massively parallel reporter assays (MPRAs) enable high-throughput testing of thousands of candidate regulatory elements simultaneously, partially addressing the scale–depth trade-off, though they retain the limitation of testing elements outside their native chromatin context [74].

## Pathway and network validation

Validating biological pathways and molecular networks is critical for confirming the results of bioinformatics analyses and understanding their role in cellular processes and disease mechanisms. Pathway-level validation poses the unique challenge of testing emergent properties of multi-gene systems, requiring approaches that go beyond single-gene perturbation.

### Gene set enrichment analysis (GSEA) and pathway analysis

Gene set enrichment analysis (GSEA) and related pathway tools identify biological processes enriched in a gene list, but enrichment results remain associative rather than causal. Interpretation should account for gene-set redundancy, background-list choice, annotation bias, and the possibility that enrichment of a pathway does not imply coordinated activation of every component. Functional validation therefore requires perturbing key nodes or directly measuring pathway output [75].

In vitro and in vivo pathway validation tests whether modulating predicted pathway components produces the expected cellular or organismal phenotype. Because signaling pathways are interconnected, pharmacological inhibition, genetic perturbation, and reporter assays should be interpreted alongside dose-response data, time-course experiments, and specificity controls.

### Validation through pharmacological inhibition and activation of pathways

Pharmacological inhibitors, agonists, siRNAs, and CRISPR-based perturbations provide complementary ways to test pathway involvement. Confidence increases when structurally distinct inhibitors, genetic perturbation, and rescue or downstream readouts produce concordant results.

## Case studies from computational to experimental validation

This section presents three focused case studies that illustrate the workflow from computational prediction through candidate prioritization and experimental validation. The examples emphasize validation logic and common failure modes rather than fixed numerical validation rates.

### Cancer gene expression: TCGA-based differential expression analysis

**Computational prediction.** In a representative cancer gene expression validation workflow, RNA-Seq data from tumor and matched normal samples are analyzed using differential expression pipelines such as DESeq2 or edgeR. Pathway enrichment analysis identifies candidate biological processes, and protein–protein interaction network analysis can prioritize genes that occupy central positions within disease-relevant modules.

**Candidate prioritization.** From the initial list of DEGs, candidates are prioritized based on consistency across independent datasets, membership in disease-relevant pathways, network centrality, clinical relevance, and novelty. This multi-criteria approach narrows a broad discovery list into a smaller set of experimentally tractable candidates.

**Validation experiments.** The validation workflow proceeds in stages of increasing stringency and cost. First, qPCR validation is performed on an independent cohort of clinical samples (e.g., paired tumor/normal tissues) to confirm differential expression at the mRNA level. Primers are designed to span exon–exon junctions to exclude genomic DNA contamination, and at least two reference genes are used for normalization. Second, Western blot analysis confirms protein-level changes for the subset of genes showing significant qPCR results. Third, immunohistochemistry (IHC) is performed on tissue microarrays to assess spatial expression patterns and confirm tumor-specific expression. Fourth, functional validation through gene knockdown (siRNA) or knockout (CRISPR) in cancer cell lines tests whether the candidate gene affects cancer-relevant phenotypes (proliferation, migration, invasion, apoptosis).

**Results and concordance.** Concordance usually decreases as validation moves from mRNA-level confirmation to protein-level assays and then to functional phenotyping. This staged attrition reflects biological regulation as well as technical differences among assays; therefore, a transcript-level result should be considered preliminary until supported by protein or functional evidence when mechanistic claims are made.

**Lessons learned.** The most common sources of discordance include: (i) batch effects between the discovery cohort (TCGA) and the validation cohort, particularly when sample processing differs; (ii) population-specific expression differences (e.g., TCGA is predominantly of European ancestry, while validation cohorts may differ); (iii) tumor heterogeneity, which can cause a gene that is differentially expressed in bulk RNA-Seq to show inconsistent expression in individual tumor samples; and (iv) post-transcriptional regulation, which decouples mRNA and protein abundance. The key lesson is that multi-level validation (mRNA, protein, functional) is essential, as concordance decreases at each successive level.

### miRNA regulation: Resolving prediction discordance through experimental validation

**Computational prediction.** Computational prediction of miRNA targets is complicated by low inter-algorithm concordance. In a representative workflow, target predictions from TargetScan, miRDB, miRanda, and DIANA-microT are compared, then filtered by consensus support, anti-correlation between miRNA and target mRNA expression, and disease-relevant pathway context.

**Candidate prioritization.** Candidates are further prioritized by consistency of expression anti-correlation, conservation of the predicted binding site, prior disease relevance, and availability of validated antibodies or reporter constructs.

**Validation experiments.** The validation strategy distinguishes between evidence for direct binding and evidence for functional regulation. Dual-luciferase reporter assays, in which the predicted target 3’ UTR is cloned downstream of firefly luciferase with Renilla luciferase as an internal control, test direct miRNA–target binding [59, 60]. Mutations in the seed-binding site serve as negative controls. qPCR and Western blot assess the effect of miRNA mimic or inhibitor transfection on endogenous target mRNA and protein levels. For targets validated at the binding and expression levels, functional experiments (proliferation, apoptosis, migration assays) test whether the miRNA–target interaction is functionally relevant in the disease context [76]. Ago-RIP (Argonaute RNA immunoprecipitation) provides transcriptome-wide evidence for physical association between the miRNA and its targets.

**Results and concordance.** Luciferase assays, endogenous expression measurements, and functional assays often produce non-identical results because direct binding, transcript abundance, protein abundance, and disease-relevant phenotype represent different evidence layers. Discordant mRNA and protein effects should be interpreted cautiously because miRNAs can regulate translation without producing large changes in mRNA abundance.

**Lessons learned.** The key insight from this workflow is that miRNA target prediction algorithms have fundamentally different strengths: TargetScan excels at predicting conserved targets with strong seed complementarity, while miRDB captures non-conserved but experimentally supported interactions. Using a consensus approach (requiring prediction by 2 algorithms) reduces the false-positive rate but increases the false-negative rate, potentially excluding biologically important non-canonical targets. Luciferase assays should always include seed-site mutant controls, as the reporter system can produce false positives through non-specific effects of miRNA overexpression. The discordance between mRNA-level and protein-level regulation highlights the importance of validating at both levels, as miRNAs can act through translational repression without substantially affecting mRNA abundance.

### PPI in Neurological Disease: From STRING network to proximity ligation

**Computational prediction.** In a representative neurological disease study (e.g., Alzheimer’s disease), disease-associated proteins identified through genome-wide association studies (GWAS) and transcriptomic analyses are used as seeds to query the STRING database for predicted protein–protein interactions [38, 77]. A first-shell interaction network (direct interactions with the seed proteins) and a second-shell network (interactions of the first-shell proteins) are constructed. Network analysis identifies interaction clusters (using MCODE or Louvain community detection), hub proteins (high degree centrality), and bottleneck proteins (high betweenness centrality). Enrichment analysis of the network clusters reveals association with specific disease-relevant pathways (e.g., neuroinflammation, synaptic signaling, protein aggregation) [78].

**Candidate prioritization.** Interactions are prioritized by STRING confidence score, evidence type supporting the score, co-expression in disease-relevant tissues, differential expression of at least one interaction partner, pathway relevance, and novelty.

**Validation experiments.** Validation employs complementary methods. Co-immunoprecipitation (Co-IP) from brain tissue lysate or neuronal cell line lysate tests whether predicted interaction partners form a complex. Proximity ligation assay (PLA) provides in situ visualization of proteins in close proximity in tissue sections or cultured neurons, adding spatial information about subcellular compartments and cell types [79]. For interactions of particular interest, FRET or split-luciferase PCA assays can provide additional evidence in living cells.

**Results and concordance.** In representative workflows, high-confidence STRING-predicted interactions should be treated as candidates for context-specific experimental validation rather than as interactions with fixed validation probabilities. Confirmation by Co-IP or PLA depends on the evidence supporting the STRING score, the biological system tested, protein expression levels, antibody performance, and whether the interaction is stable or transient [38].

**Lessons learned.** STRING scores aggregate evidence from multiple sources, including text mining, co-expression, experimental databases, and homology, and the composition of the score matters as much as its magnitude. The choice of cell type is critical because many neuronal interactions are cell-type-specific; PLA provides valuable spatial information but depends strongly on antibody availability and specificity.

## Challenges and limitations

The validation of bioinformatics predictions confronts researchers with challenges that extend well beyond the technical limitations of individual experimental methods. These challenges are best understood through the lens of the three core problems identified in this review: the specificity gap, the reproducibility–translatability tension, and the scale–depth trade-off. These challenges and corresponding validation strategies are summarized in Figure 2.

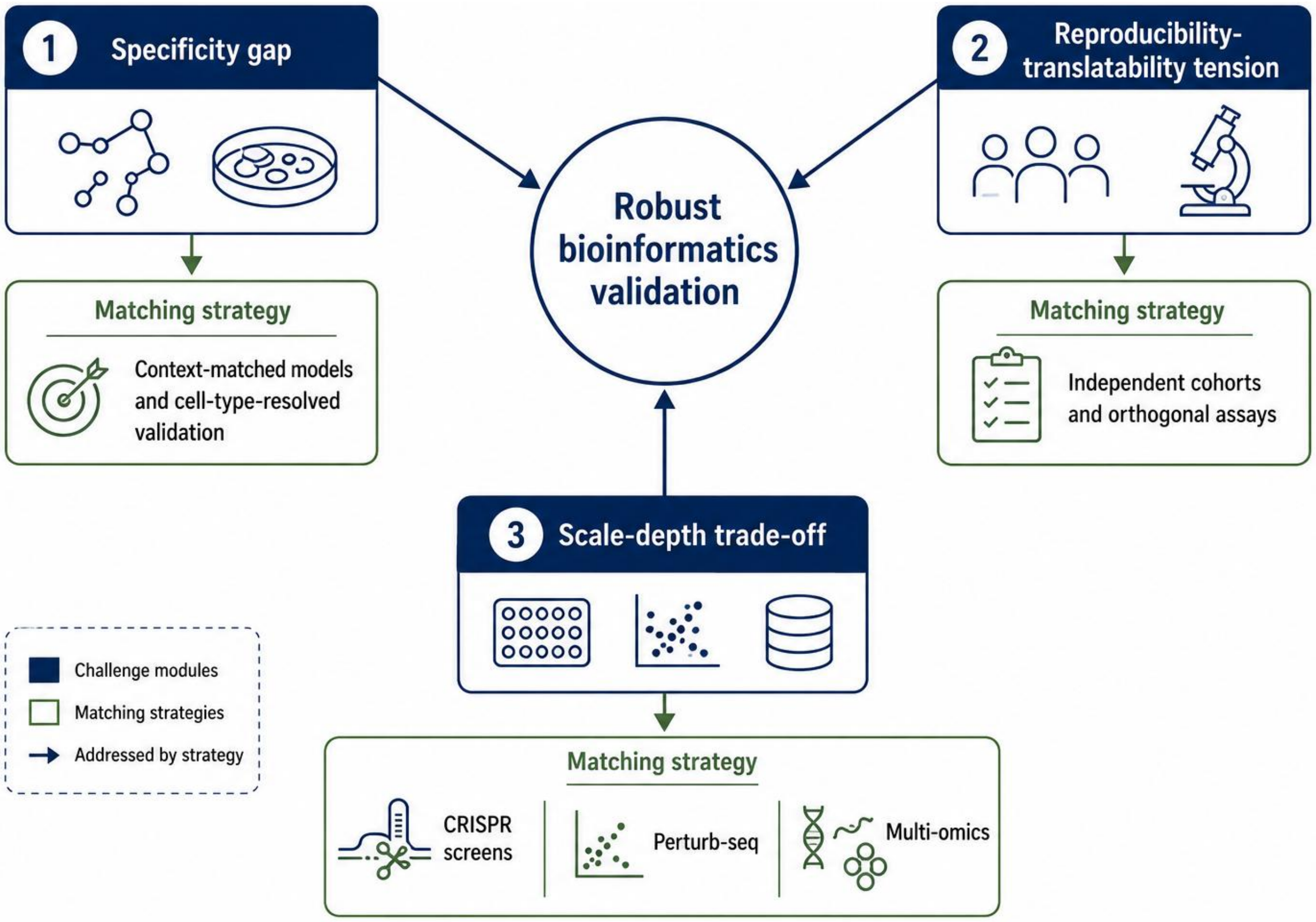

**Figure 2 Core challenges in validating bioinformatics findings and corresponding mitigation strategies.** Specificity, reproducibility, and scale constraints require context-matched models, independent replication, orthogonal assays, and scalable perturbation or multi-omics approaches.

The specificity gap arises because computational analyses operate at genome-wide scale, whereas gold-standard validation methods remain much lower throughput. For protein–protein interactions, the gap is especially clear: the complete human binary interactome is expected to contain a very large number of interactions, whereas even large-scale experimental reference maps capture only a subset. For example, the HuRI reference map reported approximately 53,000 binary protein–protein interactions, illustrating that many computationally nominated interactions still require context-specific experimental validation [80]. This quantitative imbalance means that experimentally validated interactions are likely to represent a biased subset, typically enriched for candidates with stronger prior evidence, clearer biological relevance, or greater feasibility for available assays.

The reproducibility-translatability tension arises because validation results are sensitive to reagent quality, operator choices, experimental model, cell type, tissue context, and population background. A finding confirmed in one cell line or organism may not reproduce in another because the relevant molecular interaction or pathway activity is context-dependent.

The scale-depth trade-off forces researchers to choose between broad screening and mechanistic detail. High-throughput methods can test many candidates but often provide limited mechanistic resolution, whereas low-throughput assays such as Co-IP, Western blot, and detailed phenotyping provide richer evidence for fewer candidates.

The limitations of current bioinformatics tools compound these challenges. Benchmarking studies show that different tools can produce different candidate lists from the same input data, and many databases remain incomplete, biased toward well-studied systems, or aggregated across contexts. Validation strategies should therefore account for database provenance, context specificity, and the uncertainty introduced by analytical workflow choices.

Interpreting multiple validation experiments requires distinguishing biological discordance from technical failure. Multi-omics datasets can provide cross-layer support, but integration requires careful statistical design because a gene may show mRNA-level change without a matching protein-level effect.

## Future directions and emerging technologies

The three core challenges identified in this review—the specificity gap, the reproducibility–translatability tension, and the scale–depth trade-off—are not intractable. Emerging technologies are already beginning to address each of these challenges, and their continued development will fundamentally reshape the bioinformatics validation landscape.

### CRISPR perturbation screens: Addressing the scale–depth trade-off

Perturb-seq combines pooled CRISPR perturbation with single-cell RNA sequencing, enabling massively parallel functional genomic screening with transcriptome-scale phenotypic readouts rather than simple binary hit/no-hit outcomes. Recent genome-scale Perturb-seq experiments have extended this approach to thousands of genes and millions of single cells [81]. CRISPRi and CRISPRa variants further extend this strategy to loss-of-function and gain-of-function screens, while combinatorial CRISPR approaches can test gene-gene interactions at scale.

Despite this scalability, Perturb-seq remains constrained by model choice, guide efficiency, perturbation assignment, and analysis complexity; targeted follow-up remains necessary for mechanistic interpretation.

### Single-cell technologies: Closing the specificity gap

Single-cell RNA sequencing and related technologies such as single-cell ATAC-seq, CITE-seq, and spatial transcriptomics can narrow the specificity gap by resolving cell-type-specific expression and regulatory states. These approaches are especially useful when bulk profiles mask heterogeneous responses across tumor, brain, immune, or developmental tissues.

However, algorithmic choices such as clustering resolution, trajectory inference, and batch correction can alter downstream interpretation. Spatial transcriptomics adds anatomical context, but computational predictions still require appropriate validation in the relevant cell type and tissue architecture [19].

### Artificial intelligence and machine learning: Improving reproducibility

AI and machine learning may improve validation workflows by prioritizing candidates, detecting confounders, and standardizing image or omics interpretation. Their value depends on transparent training data, external validation, and careful separation between prediction generation and validation assessment.

### Multi-omics integration: Addressing all three challenges

Multi-omics approaches integrate genomics, transcriptomics, proteomics, epigenomics, and metabolomics to provide multiple independent evidence layers for each finding. A prediction supported across data types is generally more robust than one supported by a single modality, but integration still requires context-matched samples and transparent analytical assumptions.

## Conclusion

This review has examined the computational-to-experimental validation pipeline through three recurring constraints: specificity, reproducibility, and scale. Effective validation requires deliberate combinations of orthogonal methods, context-matched models, and multi-level evidence spanning mRNA, protein, interaction, and functional readouts.

Validation strategy should be treated as a first-class design consideration rather than an afterthought. Negative or discordant validation results should also be reported because they are essential for calibrating the accuracy and practical usefulness of computational methods.

Looking ahead, CRISPR-based perturbation screens, single-cell multi-omics, and AI-assisted design are likely to increase the throughput and standardization of validation. The next challenge is to develop reporting standards that make validation outcomes comparable across laboratories and computational workflows.

### Ethical statement

Not applicable. This manuscript is a narrative review of previously published literature and does not contain any original empirical studies involving human participants, clinical tissue samples, or animal subjects conducted by any of the authors.

### Data availability

No new raw datasets were generated or computationally analyzed during the current study. All biological databases, repositories, and predictive platforms discussed in this review (including STRING, miRTarBase, and the HuRI map) are publicly accessible, and their respective configurations are fully documented within the text and reference list.

### Author contributions

Wang TY: Conceptualization, literature review, manuscript drafting, and revision. Chen SL: Literature search, data organization, and manuscript drafting. Wang YZ: Literature search, methodological synthesis, and manuscript drafting. Zhang YC: Literature review, validation framework development, and manuscript drafting. Song XY: Literature screening, table preparation, and manuscript revision. Bi ZQ: Literature screening, methodological comparison, and manuscript revision. Liu M: Literature review, case study synthesis, and manuscript revision. Niu Q: Literature review, figure/table preparation, and manuscript editing. Liu JY: Literature review, case study organization, and manuscript editing. Feng PS: Literature review, interpretation of bioinformatics validation methods, and manuscript revision. Sun XT: Literature review, emerging technology synthesis, and manuscript revision. Peng BJ: Critical review, methodological interpretation, and manuscript editing. Fei C: Literature review, interpretation of experimental validation strategies, and manuscript revision. Yan LKQ: Critical review, methodological guidance, and manuscript editing. Bao RY: Literature review, manuscript revision, and proofreading. Qin ZY: Literature review, manuscript revision, and proofreading. Jiang ZK: Supervision, conceptualization, project administration, critical revision, and final approval of the manuscript.